\documentclass[
aps,%
pre,%
a4paper,%
final,%
reprint,%
showpacs,%
superscriptaddress,%
groupedaddress,%
nofootinbib
]{revtex4-1}

\usepackage{amsmath}
\usepackage{mathrsfs}  
\usepackage{amssymb}
\usepackage{amsthm}
\usepackage{bm}

\usepackage[pdftex]{graphicx}
\graphicspath{{./figs/}}
\usepackage[pdftex,colorlinks=true,bookmarks=true,citecolor=blue,urlcolor=blue]{hyperref} 
\usepackage[caption=false]{subfig}

\newcommand{\field}[1]{\mathbb{#1}}
\newcommand{\fs}[1]{\mathsf{#1}}

\DeclareMathOperator*{\supp}{supp}
\DeclareMathOperator*{\esssup}{ess\,sup}
\DeclareMathOperator*{\Res}{Res}
\DeclareMathOperator{\diag}{diag}
\DeclareMathOperator{\Wrons}{\mathscr{W}}

\newcommand{\tp}{\intercal}

\newcommand{\ovl}[1]{\overline{#1}}

\newcommand{\bigO}[1]{\mathop{\mathscr{O}}\left(#1\right)}

\let\Re\relax
\DeclareMathOperator{\Re}{Re}
\let\Im\relax
\DeclareMathOperator{\Im}{Im}
\newcommand{\vv}[1]{\boldsymbol{#1}}
\newcommand{\vs}[1]{\boldsymbol{#1}}

\newcommand{\OP}[1]{\mathscr{#1}}

\newcommand{\wtilde}[1]{\widetilde{#1}}

\newtheorem{theorem}{Theorem}[section]

\usepackage{enumitem}

\begin{document}
\title{Exact Solution of the Zakharov-Shabat Scattering Problem for Doubly-Truncated Multi-Soliton Potentials}


\author{V.~Vaibhav}
\email{v.k.vaibhav@tudelft.nl}

\affiliation{Delft Center for Systems and Control, 
Delft University of Technology, Mekelweg 2. 2628 CD Delft, 
The Netherlands}
\date{\today}

\begin{abstract}
Recent studies have revealed that multi-soliton solutions of the nonlinear
Schr\"odinger equation, as carriers of information, offer a
promising solution to the problem of nonlinear signal distortions in fiber optic
channels. In any nonlinear Fourier transform based transmission methodology
seeking to modulate the discrete spectrum of the multi-solitons, choice
of an appropriate windowing function is an important design issue on account of
the unbounded support of such signals. Here, we
consider the rectangle function as the windowing function for the
multi-solitonic signal and provide the exact solution of 
the associated Zakharov-Shabat scattering problem for the
windowed/doubly-truncated multi-soliton
potential. This method further allows us to avoid prohibitive numerical computations normally
required in order to accurately quantify the effect of time-domain
windowing on the nonlinear Fourier spectrum of the multi-solitonic signals. The method
devised in this work also applies to general type of signals and may prove to be
a useful tool in the theoretical analysis of such systems.
\end{abstract}


\maketitle

\section*{Notations}
\label{sec:notations}
The set of non-zero positive real numbers ($\field{R}$) is denoted by
$\field{R}_+$. For any complex number $\zeta$, $\Re(\zeta)$ and $\Im(\zeta)$ refer to the real
and the imaginary parts of $\zeta$, respectively. Its complex conjugate is 
denoted by $\zeta^*$. The upper-half (lower-half) of complex plane ($\field{C}$) 
is denoted by $\field{C}_+$ ($\field{C}_-$) and it closure by $\ovl{\field{C}}_+$
($\ovl{\field{C}}_-$). Let $\sigma_0=\text{diag}(1,1)$ and the Pauli's spin
matrices are 
\[
\sigma_1=\begin{pmatrix}
0 &  1\\
1 &  0
\end{pmatrix},\quad
\sigma_2=\begin{pmatrix}
0 &  -i\\
i &  0
\end{pmatrix},\quad
\sigma_3=\begin{pmatrix}
1 &  0\\
0 & -1
\end{pmatrix}.
\]
The support of a function
$f:\Omega\rightarrow\field{R}$ in $\Omega$ is defined as $\supp
f=\ovl{\{x\in\Omega|\,f(x)\neq0\}}$. The Lebesgue spaces of complex-valued 
functions defined in $\field{R}$ are denoted by 
$\fs{L}^p$ for $1\leq p\leq\infty$ with their corresponding 
norm denoted by $\|\cdot\|_{\fs{L}^p}$ or $\|\cdot\|_p$. 

\section{Introduction}
In optical fiber communication, the propagation of optical field in a loss-less single mode
fiber under Kerr-type focusing nonlinearity is governed by the nonlinear Schr\"odinger 
equation (NSE)~\cite{HK1987,Agrawal2013} which, in its standard form, reads as
\begin{equation}\label{eq:NSE}
    i\partial_xq=\partial_t^2q+2|q|^2q,\quad(t,x)\in\field{R}\times\field{R}_+,
\end{equation}
where $q(t,x)$ is a complex valued function associated with the slowly varying
envelope of the electric field, $t$ is the retarded time and $x$ 
is position along the fiber. This equation also provides a satisfactory description 
of optical pulse propagation in the
guiding-center or path-averaged formulation~\cite{HK1990GC,HK1991GC,TBF2012} 
when more general scenarios such as
presence of fiber losses, lumped or distributed periodic amplification are
included in the mathematical model of the physical channel.

The initial value problem (IVP) corresponding to the NSE was first solved by Zakharov and
Shabat~\cite{ZS1972}, which is known to be one of the first successful implementations of 
the \emph{inverse scattering transform} (IST) method. 
\emph{Multi-solitons} or, more precisely, \emph{$K$-soliton} solutions were obtained as a
special case of this theory. The IST method was later extended to a wider
class of nonlinear evolution equations known as the Ablowitz-Kaup-Newell-Segur
(AKNS) class of integrable equations~\cite{AKNS1973,AKNS1974}. In this pioneering work,
IST was, for the first time, presented as a way of Fourier analysis for nonlinear problems
prompting researchers to coin the term \emph{nonlinear Fourier transform} (NFT)
for IST. In this terminology, any subset of the scattering data that qualifies
as the ``primordial'' scattering data~\cite{AKNS1974} is referred to as the \emph{nonlinear
Fourier spectrum}.

The fact that the energy content of $K$-solitons is not dispersed away as it
propagates along the fiber makes them promising as carriers of information 
in optical communication. These ideas 
were first explored by Hasegawa and Nyu~\cite{HN1993} who proposed encoding 
information in the eigenvalues of the $K$-soliton solutions in a framework which 
they described as the \emph{eigenvalue communication}. With the recent
breakthroughs in coherent optical communication~\cite{ILBK2008,K2016} and
growing need for increased channel capacity~\cite{MS2001,EFKW2008,EKWFG2010}, these ideas have 
been recently revived. We refer the reader to the comprehensive review 
article~\cite{TPLWFK2017} and the references therein for an overview of 
NFT-based optical communication methodologies and its potential
advantage over the conventional ones.

In this article, we focus on a particular aspect of the NFT-based transmission
methodologies which seek to modulate the discrete part of the nonlinear Fourier
spectrum using $K$-solitons as information carriers. Given that the support of
the $K$-soliton solutions is infinite, it is mandatory to
employ a windowing function~\cite{FGT2017}. The windowing function must be
such that it does not considerably alter the nonlinear Fourier spectrum of the
original signal. In this work, we consider the simplest of the windowing functions, the
\emph{rectangle function}. It is shown that the resulting scattering problem for
the ``windowed'' or the doubly-truncated $K$-soliton solutions is exactly solvable. The idea is to
express the Jost solutions of the windowed potential in terms of the Jost solutions
of the original potential. Such an approach has already appeared in the work of
Lamb~\cite{Lamb1980} where the scattering problem for a potential truncated from
one side is solved exactly using the Jost solutions of the original potential. In
particular, the observation that truncated $K$-soliton has rational reflection coefficient 
has been used to devise exact techniques for IST~\cite{RM1992,RS1994,SK2008}.
Adapting Lamb's approach, it is further shown that, in the case of truncation from both
sides, one can set up a Riemann-Hilbert (RH) problem to obtain the Jost
solutions of the doubly-truncated potential. It must be noted that this method applies to general
potentials; however, for $K$-solitons, evaluation of certain integrals become
a trivial task and the solution of the RH-problem can be obtained in a closed
form. In particular, the method of \emph{Darboux transformation} (DT) for computing
$K$-solitons provides an adequate representation of the Jost solutions in terms
of the so called \emph{Darboux matrix} which, as a function of the spectral
parameter, has a rational structure facilitating the solution of the
RH-problem. This representation further enables us to obtain precise estimates for the 
effective temporal support as well as spectral width of the $K$-soliton pulses. The 
rational structure of the aforementioned Darboux matrix has also been
recently exploited to develop fast numerical algorithms for DT~\cite{V2017INFT1} 
and IST~\cite{V2017arXivFastINFT}.

\section{Direct Scattering: Doubly-Truncated Potential}
\label{sec:akns-sys}
The NFT of a given complex-valued signal $q(t)$ is introduced 
via the associated \emph{Zakharov-Shabat scattering problem} (or ZS-problem in
short)~\cite{ZS1972} 
which can be stated as follows:
Let $\zeta\in\field{R}$ and $\vv{v}=(v_1,v_2)^{\tp}\in\field{C}^2$, then 
\begin{equation}\label{eq:zs-prob}
\vv{v}_t = -i\zeta\sigma_3\vv{v}+U\vv{v},
\end{equation}
where the matrix elements of $U$ are $U_{11}=U_{22}=0$ and
$U_{12}=q(t)=-U_{21}^*=-r^*(t)$. Here, $q(t)$ is identified as the 
\emph{scattering potential}. Henceforth, we closely 
follow the formalism developed in~\cite{AKNS1974,AS1981}. We 
assume that the Jost solutions of the \emph{first kind}, denoted
by $\vs{\psi}(t;\zeta)$ and $\overline{\vs{\psi}}(t;\zeta)$, which are the linearly
independent solutions of~\eqref{eq:zs-prob}, are known. These solutions are
characterized by the following asymptotic 
behavior as $t\rightarrow\infty$:
$\vs{\psi}(t;\zeta)e^{-i\zeta t}\rightarrow(0,1)^{\tp}$ and 
$\overline{\vs{\psi}}(t;\zeta)e^{i\zeta t}\rightarrow (1,0)^{\tp}$. 
We also assume that the Jost solutions of the \emph{second kind}, denoted by
$\vs{\phi}(t,\zeta)$ and $\overline{\vs{\phi}}(t,\zeta)$, which are also linearly 
independent solutions of~\eqref{eq:zs-prob} are known. These solutions are
characterized by the following asymptotic behavior as $t\rightarrow-\infty$: 
$\vs{\phi}(t;\zeta)e^{i\zeta t}\rightarrow(1,0)^{\tp}$ and 
$\overline{\vs{\phi}}(t;\zeta)e^{-i\zeta t}\rightarrow(0,-1)^{\tp}$.
The scattering coefficients corresponding to $q(t)$ can be written in terms of the
Jost solutions by using the Wronskian relations~\cite{AKNS1974}
\begin{equation}\label{eq:wrons-scoeff}
\begin{split}
&a(\zeta)= \Wrons\left(\vs{\phi},{\vs{\psi}}\right),\quad 
 b(\zeta)= \Wrons\left(\ovl{\vs{\psi}},\vs{\phi}\right),\\
&\ovl{a}(\zeta)=\Wrons\left(\ovl{\vs{\phi}},\overline{\vs{\psi}}\right),\quad
\ovl{b}(\zeta) =\Wrons\left(\ovl{\vs{\phi}},{\vs{\psi}}\right).
\end{split}
\end{equation}
Furthermore, the symmetry properties, 
$\ovl{\vv{\psi}}(t;\zeta)=i\sigma_2\vv{\psi}^*(t;\zeta^*)$ and 
$\ovl{\vv{\phi}}(t;\zeta)=i\sigma_2\vv{\phi}^*(t;\zeta^*)$
yield the relations $\ovl{a}(\zeta)=a^*(\zeta^*)$ and $\ovl{b}(\zeta)=b^*(\zeta^*)$.

Here, we assume that the nonlinear Fourier spectrum of the signal $q(t)$ is as
follows: The discrete spectrum consists 
of the eigenvalues $\zeta_k\in\field{C}_+$ and the norming constants $b_k$. For 
convenience, let the discrete spectrum be denoted by the set
\begin{equation}
\mathfrak{S}_K=\{(\zeta_k,\,b_k)\in\field{C}^2,\,k=1,2,\ldots,K\}.
\end{equation}
The continuous spectrum, also referred to as the \emph{reflection coefficient}, is given 
by $\rho(\xi)={b(\xi)}/{a(\xi)}$ for $\xi\in\field{R}$.

In this article, we consider windowing using the rectangle function supported in
$[-T_-, T_+]$ where $T_-, T_+>0$. Define the left-sided signal $q^{(-)}(t;T_+) =
q(t)\theta(T_+-t)$, where $\theta(t)$ is the Heaviside step function so that the 
windowed signal is $q^{(\sqcap)}(t;T_-,T_+)=q^{(-)}(t;T_+)\theta(t+T_-)$. Here,
our objective is to solve the ZS-problem corresponding to the windowed potential 
$q^{(\sqcap)}(t;T_-,T_+)$. To this end, we first derive the Jost solutions for 
the left-sided signal $q^{(-)}(t;T_+)$ supported in $(-\infty, T_+]$. Starting 
from the Jost solution of the second kind, it is straightforward to verify that 
(for $\zeta\in\ovl{\field{C}}_+$)
\begin{equation}
\vs{\phi}^{(-)}(t;\zeta)=
\begin{cases}
\vs{\phi}(t;\zeta), & t\leq T_+,\\
e^{-i\sigma_3\zeta(t-T_+)}\vs{\phi}(T_+;\zeta), & t>T_+.
\end{cases}
\end{equation}
For $t\geq T_+$, the potential is identically zero so that 
$\vs{\psi}^{(-)}(t;\zeta)=(0,1)^{\tp}e^{i\zeta t},\,\zeta\in\ovl{\field{C}}_+$. 
Now, using the Wronskian relations~\eqref{eq:wrons-scoeff}, the 
scattering coefficients, for $\zeta\in\ovl{\field{C}}_+$, work out to be
$a^{(-)}(\zeta) = \phi_1(T_+;\zeta)e^{i\zeta T_+}$ and $b^{(-)}(\zeta) =
\phi_2(T_+;\zeta)e^{-i\zeta T_+}$.
Next, our aim is to obtain $\vs{\psi}^{(-)}(t;\zeta)$ for $t<T_+$ for
$\zeta\in\field{C}_+$. On the real axis, i.e. $\xi\in\field{R}$, one can obtain 
$\vs{\psi}^{(-)}(t;\xi)$ using the linear independence 
of $\vs{\phi}^{(-)}(t;\xi)$ and $\ovl{\vs{\phi}^{(-)}}(t;\xi)$:
\begin{equation}
\vs{\psi}^{(-)}(t;\xi)=
-a^{(-)}(\xi)\ovl{\vs{\phi}^{(-)}}(t;\xi)\\
+\ovl{b^{(-)}}(\xi)\vs{\phi}^{(-)}(t;\xi).
\end{equation}
Evidently, the expression in the RHS above cannot be analytically continued into the upper-half of
the complex plane. In order to circumvent this limitation, we adopt a different
approach in the following. Let us consider the relation
\begin{equation}
\vs{\phi}^{(-)}(t;\zeta)
= a^{(-)}(\zeta)\ovl{\vs{\psi}^{(-)}}(t;\zeta)
+b^{(-)}(\zeta)\vs{\psi}^{(-)}(t;\zeta),
\end{equation}
for $\zeta\in\field{R}$. Let $a^{(-)}(\zeta)$ have $K'$ simple zeros in
$\field{C}_+$ denoted by $\zeta^{(-)}_k,k=1,2,\ldots,K'$. Following~\cite{ZS1972}, we 
set up a Riemann-Hilbert (RH) problem for the sectionally analytic
function $\vv{F}(\zeta)\equiv\vv{F}(\zeta;t)$ with simple poles in $\field{C}_+$ defined by
\begin{equation}
\vv{F}(\zeta;t)=
\begin{cases}
[a^{(-)}(\zeta)]^{-1}\vs{\phi}^{(-)}(t;\zeta)e^{i\zeta t},&\zeta\in\field{C}_+,\\
\ovl{\vs{\psi}^{(-)}}(t;\zeta)e^{i\zeta t},&\zeta\in\field{C}_-,
\end{cases}
\end{equation}
with the jump condition given by
\begin{equation}
\vv{F}(\xi+i0)-\vv{F}(\xi-i0) =
\rho^{(-)}(\xi)\vs{\psi}^{(-)}(t;\xi)e^{i\xi t},
\end{equation}
for $\xi\in\field{R}$ where $\rho^{(-)}(\xi)=b^{(-)}(\xi)/a^{(-)}(\xi)$. 
The solution of the RH-problem can be stated as
\begin{multline}
\label{eq:soln-RH}
\vv{F}(\zeta)=
\begin{pmatrix}
1\\
0
\end{pmatrix}
+\sum_{k=1}^{K'}\frac{e^{i\zeta_k t}}{\left(\zeta-\zeta^{(-)}_k\right)}
\frac{\vs{\phi}^{(-)}(t;\zeta^{(-)}_k)}{\dot{a}^{(-)}(\zeta^{(-)}_k)}\\
+\frac{1}{2\pi i}\int_{-\infty}^{\infty}{\rho^{(-)}(\xi)\vs{\psi}^{(-)}(t;\xi)
e^{i\xi t}}\frac{d\xi}{\xi-\zeta}.
\end{multline}
Note that for $t<T_+$ and
$\xi\in\field{R}$, we have
\begin{multline}\label{eq:RH-integrand}
\rho^{(-)}(\xi)\vs{\psi}^{(-)}(t;\xi)e^{i\xi t}=
-b^{(-)}(\xi)\ovl{\vs{\phi}^{(-)}}(t;\xi)e^{i\xi t}\\
+\frac{|b^{(-)}(\xi)|^2}{a^{(-)}(\xi)}\vs{\phi}^{(-)}(t;\xi)e^{i\xi t}.
\end{multline}
The RHS of the above equation is known for all $\xi\in\field{R}$; therefore,
$\vv{F}(\zeta)$ can be obtained explicitly provided that the integral
in~\eqref{eq:soln-RH} can be computed exactly. This yields 
$\vs{\psi}^{(-)}(t;\zeta)$ for $t<T_+$ and $\zeta\in\field{C}_+,$ using the
symmetry properties. 

Next, the windowed potential $q^{(\sqcap)}(t;T_-,T_+)$ is obtained as a result of 
truncation of $q^{(-)}(t;T_+)$ from left. Let
the Jost solutions for this potential be $\vs{\psi}^{(\sqcap)}(t;\zeta)$ (first kind) and 
$\vs{\phi}^{(\sqcap)}(t;\zeta)$ (second kind). Let the scattering coefficients be denoted by
$a^{(\sqcap)}(\zeta)$ and $b^{(\sqcap)}(\zeta)$.
In the following, our aim would be to obtain an
expression for these Jost solutions in terms of the Jost solutions of
$q^{(-)}(t;T_+)$. Evidently, 
\begin{equation}
\vs{\psi}^{(\sqcap)}=
\begin{cases}
e^{-i\sigma_3\zeta(t+T_-)}\vs{\psi}^{(-)}(-T_-;\zeta), & t<-T_-\\
\vs{\psi}^{(-)}(t;\zeta), & t\geq -T_-,\\
\end{cases}
\end{equation}
and $\vs{\phi}^{(\sqcap)}(t;\zeta)=(1,0)^{\tp}e^{-i\zeta t}$ for $t<-T_-$ where 
$\zeta\in\ovl{\field{C}}_+$. Now using the Wronskian relations, we have 
\begin{equation}
\begin{split}
a^{(\sqcap)}(\zeta)
&= \psi_2^{(\sqcap)}(-T_-;\zeta)e^{i\zeta T_-}
=F_1^*(\zeta^*;-T_-),\\
\ovl{b^{(\sqcap)}}(\zeta)
&= \psi_1^{(\sqcap)}(-T_-;\zeta)e^{-i\zeta T_-}\\
&=-F^*_2(\zeta^*;-T_-)e^{-2i\zeta T_-},
\end{split}
\end{equation}
for $\zeta\in\ovl{\field{C}}_+$. This implies
${b^{(\sqcap)}}(\zeta)=-F_2(\zeta;-T_-)e^{2i\zeta T_-}$ for
$\zeta\in\ovl{\field{C}}_-$. Note that the functional form obtained for the
scattering coefficients $a^{(\sqcap)}(\zeta)$ and  $b^{(\sqcap)}(\zeta)$ must
hold for all $\zeta\in\field{C}$ if they hold true in any of the half-planes.

The expression in~\eqref{eq:soln-RH} appears to indicates
that zeros of $a^{(-)}(\zeta)$ are required in order to compute
$\vv{F}(\zeta)$; however, a closer look at this expression shows that it is not the 
case: Consider, for $t<T_+$,
\begin{multline*}
\frac{1}{2\pi i}\int_{-\infty}^{\infty}
\frac{|b^{(-)}(\xi)|^2}{a^{(-)}(\xi)}\vs{\phi}^{(-)}(t;\xi)e^{i\xi
t}\frac{d\xi}{\xi-\zeta}\\
=\frac{1}{2\pi i}\int_{-\infty}^{\infty}
\left[\frac{1}{a^{(-)}(\xi)}-a^{(-)*}(\xi)\right]
\vs{\phi}^{(-)}(t;\xi)e^{i\xi t}\frac{d\xi}{\xi-\zeta}.
\end{multline*}
Given that $1/{a^{(-)}(\xi)}$ is holomorphic in $\field{C}_+$ with isolated
poles at $\zeta^{(-)}_k$, this part of the integrand can be computed easily by
completing the contour in $\field{C}_+$ so that
\begin{widetext}
\begin{equation*}
\frac{1}{2\pi i}\int_{-\infty}^{\infty}
\frac{\vs{\phi}^{(-)}(t;\xi)}{a^{(-)}(\xi)}\frac{e^{i\xi t}d\xi}{\xi-\zeta}
+\sum_{k=1}^{K'}\frac{e^{i\zeta_k t}}{\left(\zeta-\zeta^{(-)}_k\right)}
\frac{\vs{\phi}^{(-)}(t;\zeta^{(-)}_k)}{\dot{a}^{(-)}(\zeta^{(-)}_k)}
=\begin{cases}
[a^{(-)}(\zeta)]^{-1}{\vs{\phi}^{(-)}(t;\zeta)}e^{i\zeta t},
&\zeta\in\field{C}_+,\\
0,&\zeta\in\field{C}_-
\end{cases}
\end{equation*}
\begin{equation}
\Rightarrow\vv{F}(\zeta)=
\begin{pmatrix}
1\\
0
\end{pmatrix}
-\frac{1}{2\pi i}\int_{-\infty}^{\infty}\biggl[b^{(-)}(\xi)\ovl{\vs{\phi}^{(-)}}(t;\xi)
+a^{(-)*}(\xi)\vs{\phi}^{(-)}(t;\xi)\biggl]e^{i\xi t}\frac{d\xi}{\xi-\zeta},\quad\zeta\in\field{C}_-.
\end{equation}
\end{widetext}



\section{Doubly-Truncated Multi-Soliton Potential}
Having obtained the general recipe above, we now turn to the case of
$K$-soliton potentials. The $K$-soliton potentials along with their Jost 
solutions can be computed quite easily using the 
\emph{Darboux transformation} (DT)~\cite{NM1984,Lin1990,GHZ2005}. In this discussion,
we use the DT procedure described in~\cite{NM1984}. Let $\mathfrak{S}_K$ 
be the discrete spectrum of a $K$-soliton potential. Define the matrix form of the Jost 
solutions as ${v}(t;\zeta) = (\vs{\phi},\vs{\psi})$. The seed solution here corresponds to 
the \emph{null} potential; therefore, 
$v_0(t;\zeta)=e^{-i\sigma_3\zeta t}$.
The augmented matrix Jost solution ${v}_K(t;\zeta)$ can be obtained from the 
seed solution $v_0(t;\zeta)$ using the Darboux matrix as 
${v}_K(t;\zeta)=\mu_{K}(\zeta)D_{K}(t;\zeta,\mathfrak{S}_K)v_0(t;\zeta)$ for 
$\zeta\in\ovl{\field{C}}_+$
where the Darboux matrix is written as 
$D_{K}(t;\zeta,\mathfrak{S}_K)=\sum_{k=0}^{K}D_k^{(K)}(t;\mathfrak{S}_K)\zeta^k$
where the coefficient matrices are such that $D^{(K)}_{K} = \sigma_0$ 
and 
\begin{equation}\label{eq:DT-mat-coeffs}
D_k^{(K)} = 
\begin{pmatrix}
 {d}^{(k, K)}_{0} & {d}^{(k, K)}_{1}\\
-{d}^{(k, K)*}_{1} & {d}^{(k, K)*}_{0}
\end{pmatrix},
\end{equation}
where $k=0,1,\ldots,K-1$. Also, let us recall 
$a_{K}(\zeta)=\prod_{k=1}^{K}{(\zeta-{\zeta}_k)}{(\zeta-{\zeta}^*_k)^{-1}}$
and $(\mu_{K})^{-1}=\prod_{k=1}^{K}{(\zeta-{\zeta}^*_k)}$~\cite{V2017INFT1}.
For $\zeta\in\ovl{\field{C}}_+$, it is known that~\cite{AKNS1974} 
\begin{equation}
{v}_Ke^{i\sigma_3\zeta t}=
\begin{pmatrix}
 1+\frac{1}{2i\zeta}\mathcal{E}^{(-)}& \frac{1}{2i\zeta}q(t)\\
-\frac{1}{2i\zeta}r(t)& 1+\frac{1}{2i\zeta}\mathcal{E}^{(+)}
\end{pmatrix}+\bigO{\frac{1}{\zeta^2}},
\end{equation}
where 
\begin{equation}
\begin{split}
&\mathcal{E}^{(-)}(t)=\int^t_{-\infty}|q(s)|^2ds,\\
&\mathcal{E}^{(+)}(t)=\int_t^{\infty}|q(s)|^2ds.
\end{split}
\end{equation}
This allows us to conclude that $q(t) = 2i{d}^{(K-1, K)}_{1}$ and
\begin{equation}
\begin{split}
&\mathcal{E}^{(-)} = 2i{d}^{(K-1, K)}_{0}+2i\sum_{k=1}^K\zeta_k^*,\\
&\mathcal{E}^{(+)} = 2i{d}^{(K-1, K)*}_{0}+2i\sum_{k=1}^K\zeta_k^*.
\end{split}
\end{equation}
The energy in the tails (i.e., the part of the signal
outside $[-T_-,T_+]$) is
$\epsilon_{\text{tails}}\|q\|_2^2=\mathcal{E}_-(-T_-)+\mathcal{E}_+(T_+)$ where
$\|q\|_2^2=4\sum_{k=1}^{K}\Im{\zeta_k}$.

\begin{figure*}[!t]
\centering
\includegraphics[scale=0.8]{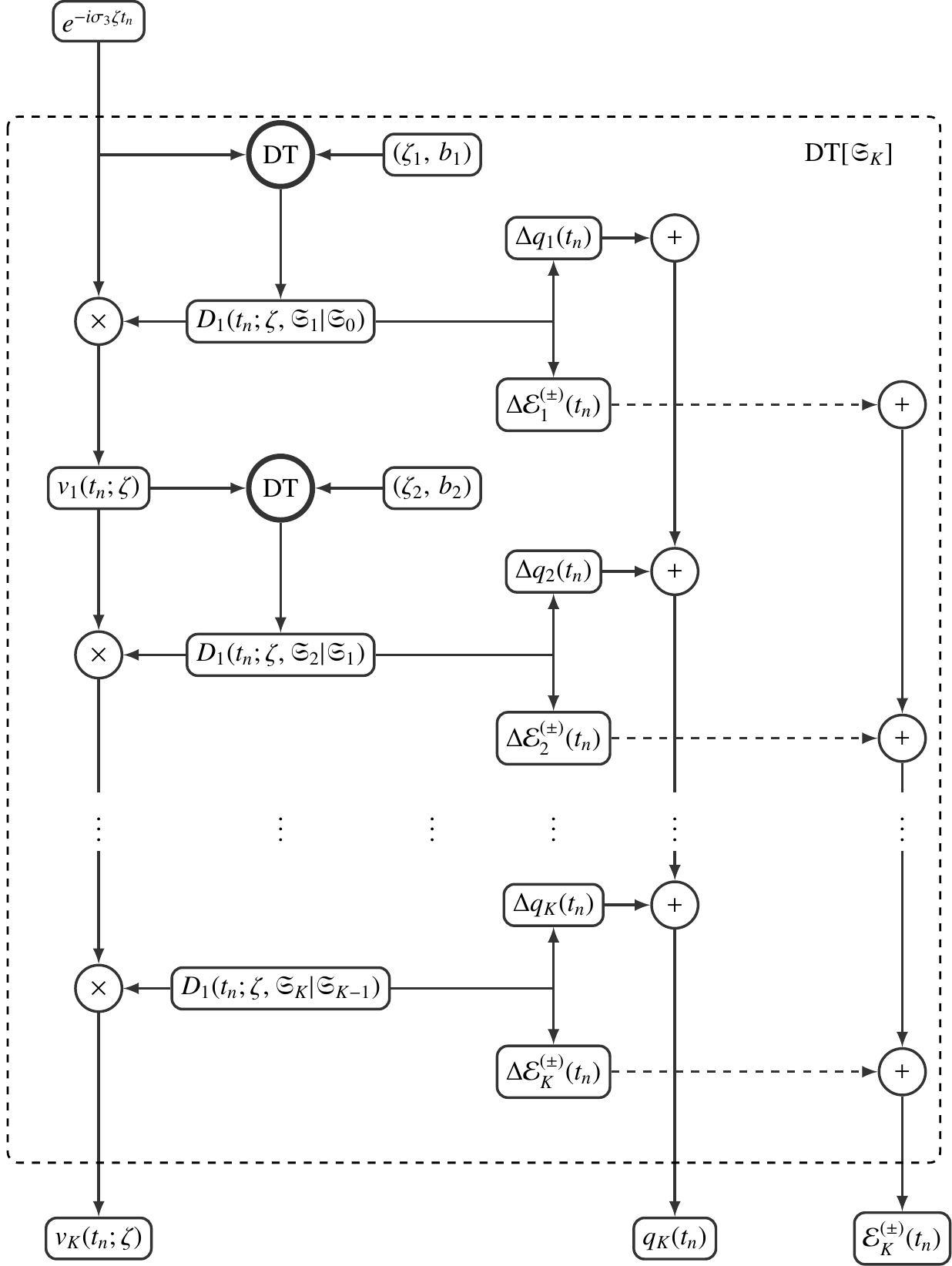}
\caption{\label{fig:DT-block}%
The figure shows the schematic of the 
Darboux transformation for a given discrete spectrum $\mathfrak{S}_K$ at 
the grid point $t_n$. The input is the seed Jost
solution, $v_0(t_n;\zeta)=e^{-i\sigma_3\zeta t_n}$. Here, $\Delta q_j(t_n) = q_j(t_n)
- q_{j-1}(t_n)$ and $\Delta \mathcal{E}^{(\pm)}_j(t_n) =
\mathcal{E}_j^{(\pm)}(t_n)-\mathcal{E}_{j-1}^{(\pm)}(t_n)$.}
\end{figure*}
The Darboux transformation can be implemented as a recursive scheme~\cite{GHZ2005}. 
Noting that the seed potential is a null potential, its discrete spectrum
is empty. Let us define the successive discrete spectra
$\emptyset=\mathfrak{S}_0\subset\mathfrak{S}_1\subset\mathfrak{S}_2
\subset\ldots\subset\mathfrak{S}_K$ such that 
${\mathfrak{S}}_j=\{(\zeta_j,b_j)\}\cup{\mathfrak{S}}_{j-1}$ for
$j=1,2,\ldots,K$ where $(\zeta_j,b_j)$ are distinct elements of 
$\mathfrak{S}_K$. The Darboux matrix of degree $K>1$ can be factorized into
Darboux matrices of degree one as
\begin{multline}
D_K(t;\zeta,\mathfrak{S}_K|\mathfrak{S}_0)
=D_1(t;\zeta,\mathfrak{S}_{K}|\mathfrak{S}_{K-1})\\\times
D_1(t;\zeta,\mathfrak{S}_{K-1}|\mathfrak{S}_{K-2})\times
\ldots\times D_1(t;\zeta,\mathfrak{S}_1|\mathfrak{S}_0),
\end{multline}
where $D_1(t;\zeta,\mathfrak{S}_{j}|\mathfrak{S}_{j-1}),\,j=1,\ldots,K$ are 
the successive Darboux matrices of degree 
one with the convention that 
$(\zeta_{j},b_{j})=\mathfrak{S}_{j}\cap\mathfrak{S}_{j-1}$ is the bound
state being added to the seed potential whose discrete spectra is
$\mathfrak{S}_{j-1}$. The resulting scheme is depicted in
Fig.~\ref{fig:DT-block}. Note that the Darboux matrices of degree one can be
stated as
\begin{multline}
D_1(t;\zeta,\mathfrak{S}_{j}|\mathfrak{S}_{j-1})= \zeta \sigma_0-\\
\begin{pmatrix}
\frac{|\beta_{j-1}|^2\zeta_j+\zeta_j^*}{1+|\beta_{j-1}|^2} 
&\frac{(\zeta_j-\zeta_j^*)\beta_{j-1}}{1+|\beta_{j-1}|^2}\\
\frac{(\zeta_j-\zeta_j^*)\beta^*_{j-1}}{1+|\beta_{j-1}|^2}
&\frac{\zeta_j+\zeta_j^*|\beta_{j-1}|^2}{1+|\beta_{j-1}|^2}
\end{pmatrix},
\end{multline}
and
\begin{equation}
\beta_{j-1}(t;\zeta_j, b_j) =
\frac{\phi_1^{(j-1)}(t;\zeta_j)-b_{j}\psi_1^{(j-1)}(t;\zeta_j)}
{\phi_2^{(j-1)}(t;\zeta_j) - b_{j}\psi_2^{(j-1)}(t;\zeta_j)},
\end{equation}
for $(\zeta_j,b_j)\in\mathfrak{S}_K$ and the successive Jost solutions, 
${v}_{j} = (\vs{\phi}_{j},\vs{\psi}_{j})$, needed in this ratio are computed as
\begin{equation}
    {v}_j(t;\zeta)=\frac{1}{(\zeta-\zeta^*_j)}D_{1}(t;\zeta,\mathfrak{S}_j|\mathfrak{S}_{j-1})
    v_{j-1}(t;\zeta).
\end{equation}
The
potential is given by 
\begin{equation}\label{DT-iter-pot}
q_j = q_{j-1} -
2i\frac{(\zeta_j-\zeta_j^*)\beta_{j-1}}{1+|\beta_{j-1}|^2}.
\end{equation}
and
\begin{equation}
\begin{split}
\mathcal{E}^{(-)}_j &=\mathcal{E}^{(-)}_{j-1}
+\frac{4\Im(\zeta_j)}{1+|\beta_{j-1}|^{-2}},\\
\mathcal{E}^{(+)}_j &=\mathcal{E}^{(+)}_{j-1}
+\frac{4\Im(\zeta_j)}{1+|\beta_{j-1}|^{2}}.
\end{split}
\end{equation}
Finally, let us observe that the computation of the Darboux matrix coefficients 
can be done in $\bigO{K^2}$ operations. With $N$ samples of $q(t)$
over $[-T_-,T_+]$, the complexity of computing $\epsilon_{\text{tails}}$ using
the trapezoidal rule (TR) of integration is $\bigO{K^2N}$ which can be contrasted
with the method proposed here using DT-coefficients which affords a complexity
of $\bigO{K^2}$ yielding an accuracy up to the machine precision\footnote{For a given 
discrete spectrum and $\epsilon_{\text{tails}}$, one can also determine 
$T_{\pm}$ using a binary search method (see~\cite{V2017NS_PREs1}).}. Further, if one
attempts to study the effect of propagation of the pulse over $M$ points along
the fiber, our method affords a complexity of $\bigO{K^2M}$ as opposed to
$\bigO{K^2NM}$ of any numerical method.

Next, the scattering coefficients 
corresponding to the truncated $K$-soliton potential $q^{(-)}(t;T_+)$ work out
to be $a^{(-)}(\zeta) = \mu_{K}(\zeta)[D_K(T_+;\zeta,\mathfrak{S}_K)]_{11}$ and 
$b^{(-)}(\zeta) =
\mu_{K}(\zeta)[D_K(T_+;\zeta,\mathfrak{S}_K)]_{21}e^{-2i\zeta T_+}$.
In the following, we suppress the dependence on $\mathfrak{S}_K$ for the sake of
brevity and proceed to construct the Jost solution $\psi^{(-)}$. Firstly, in order to 
facilitate the solution of the RH-problem introduced above, we intend to 
compute the terms in~\eqref{eq:soln-RH} exactly by exploiting the rational structure of 
the Darboux matrix. To this end, let us note that the expression 
in~\eqref{eq:RH-integrand}, for $t<T_+$, can be written as
\begin{equation}\label{eq:pq-rational}
\rho^{(-)}(\xi)\vs{\psi}_K^{(-)}(t;\xi)e^{i\xi t}=
\vv{P}(\xi;t)e^{2i\xi (t-T_+)}+\vv{Q}(\xi;t),
\end{equation}
where $\vv{P}(\xi;t)$ and $\vv{Q}(\xi;t)$ are vector-valued rational 
functions of $\xi$. These functions can be explicitly stated in terms of the
Darboux matrix elements as follows:
\begin{align}
&{\vv{P}}= -{|\mu_K(\xi)|^2}[D_K(T_+;\xi)]_{21}
\begin{pmatrix}
[D^*_K(t;\xi)]_{21}\\
-[D^*_K(t;\xi)]_{11}
\end{pmatrix},\\
&{\vv{Q}} =
{|\mu_K(\xi)|^2}\frac{|[D_K(T_+;\xi)]_{21}|^2}{[D_K(T_+;\xi)]_{11}}
\begin{pmatrix}
[D_K(t;\xi)]_{11}\\
[D_K(t;\xi)]_{21}
\end{pmatrix}.
\end{align}
From these expressions, it follows that the poles of the rational function 
$\vv{P}(\xi;t)$ are $\zeta_k$ and $\zeta_k^*$ while the poles of the rational 
function $\vv{Q}(\xi;t)$ are $\zeta_k$, $\zeta_k^*$
and zeros of $a^{(-)}(\zeta)$. For the sake of convenience, let us introduce the
residues: $\omega_j(\zeta_k)=\Res\left(Q_j(\xi;t);\zeta_k\right)$, 
$\omega_j(\zeta^{(-)}_k)=\Res\left(Q_j(\xi;t);\zeta^{(-)}_k\right)$, 
$\pi_j(\zeta_k)=\Res\left(P_j(\xi;t);\zeta_k\right)$ and 
$\pi_j(\zeta^*_k)=\Res\left(P_j(\xi;t);\zeta^*_k\right)$ for $j=1,2$. 
For $\zeta\in\field{C}_-$, define
\begin{align}
\mathcal{I}_j(\zeta;t)
&=\lim_{\kappa\rightarrow\infty}\frac{1}{2\pi i}\oint_{\Gamma_{\kappa}}
\frac{d\xi}{\xi-\zeta}{{P}_j(\xi;t)}e^{2i\xi(t-T_+)},\\
\mathcal{J}_j(\zeta;t)
&=\lim_{\kappa\rightarrow\infty}
\frac{1}{2\pi i}\oint_{\Gamma_{\kappa}}
\frac{{Q}_j(\xi;t)d\xi}{\xi-\zeta}\nonumber\\
&\quad+\sum_{k=1}^{K'}\frac{\omega_j(\zeta^{(-)}_k)}{\zeta-\zeta_k^{(-)}}
=-\sum_{k=1}^{K}\frac{\omega_j(\zeta_k)}{\zeta-\zeta_k}\label{eq:J-func},
\end{align}
where 
$\Gamma_{\kappa}$ denotes the contour comprising the segment $[-\kappa,\kappa]$ 
($\kappa>0$) and a semicircular arc with radius $\kappa$ oriented negatively in $\field{C}_-$ 
and $j=1,2$. Observing,
\begin{equation}
P_j(\xi;t)=
\sum_{k=1}^K\left[\frac{\pi_j(\zeta_k)}{\xi-\zeta_k}
+\frac{\pi_j(\zeta_k^*)}{\xi-\zeta_k^*}\right],
\end{equation}
the integrals $\mathcal{I}_j$ work out to be
\begin{equation}\label{eq:I-func}
\begin{split}
\mathcal{I}_j(\zeta;t)
&=-\sum_{k=1}^K\frac{\pi_j(\zeta_k)}{\zeta-\zeta_k}e^{2i\zeta (t-T_+)}\\
&\quad-\sum_{k=1}^K\frac{\pi_j(\zeta_k^*)}{\zeta-\zeta_k^*}
\left[e^{2i\zeta (t-T_+)}-e^{2i\zeta_k^* (t-T_+)}\right].
\end{split}
\end{equation}
This allows us to write 
$\vv{F}(\zeta)=(1,0)^{\tp}+\vv{\mathcal{I}}(\zeta;t)+\vv{\mathcal{J}}(\zeta;t)$ 
where we have used the fact that the second term in the RHS
of~\eqref{eq:soln-RH} is given by
\begin{equation*}
\sum_{k=1}^{K'}\frac{1}{\zeta-\zeta^{(-)}_k}
\begin{pmatrix}
\frac{[D_K(t;\zeta^{(-)}_k)]_{11}}{[\dot{D}_K(T_+;\zeta^{(-)}_k)]_{11}}\\
\frac{[D_K(t;\zeta^{(-)}_k)]_{21}}{[\dot{D}_K(T_+;\zeta^{(-)}_k)]_{11}}
\end{pmatrix}=\sum_{k=1}^{K'}\frac{\vs{\omega}(\zeta^{(-)}_k)}{\zeta-\zeta_k^{(-)}},
\end{equation*}
with $\vs{\omega}=(\omega_1,\omega_2)^{\tp}$. Next, let us show that the poles of 
$\vv{F}(\zeta)$ at $\zeta_k$ (as well as at $\zeta_k^*$) are removable. First let us observe that
\[
\begin{pmatrix}
\phi_1(t;\zeta)e^{i\zeta t} &\psi_1(t;\zeta)e^{-i\zeta t}\\
\phi_2(t;\zeta)e^{i\zeta t} &\psi_2(t;\zeta)e^{-i\zeta t}
\end{pmatrix} 
=D_K(t;\zeta_k).
\]
Using the symmetry relations for the Darboux 
matrix, let us also observe that 
\begin{align*}
&-\frac{[D^*_K(t;\zeta^*_k)]_{21}}{[D_K(t;\zeta_k)]_{11}}e^{2i\zeta_kt}
+\frac{[D^*_K(T_+;\zeta^*_k)]_{21}}{[D_K(T_+;\zeta_k)]_{11}}e^{2i\zeta_kT_+}\\
&=\frac{[D_K(t;\zeta_k)]_{12}}{[D_K(t;\zeta_k)]_{11}}e^{2i\zeta_kt}
-\frac{[D_K(T_+;\zeta_k)]_{12}}{[D_K(T_+;\zeta_k)]_{11}}e^{2i\zeta_kT_+}\\
&=\frac{\psi_1(t;\zeta_k)}{\phi_1(t;\zeta_k)}-
\frac{\psi_1(T_+;\zeta_k)}{\phi_1(T_+;\zeta_k)}\equiv0,
\end{align*}
on account of the property of the norming
constant $b_k$. Similarly,
\begin{align*}
&\frac{[D^*_K(t;\zeta^*_k)]_{11}}{[D_K(t;\zeta_k)]_{21}}e^{2i\zeta_kt}
+\frac{[D^*_K(T_+;\zeta^*_k)]_{21}}{[D_K(T_+;\zeta_k)]_{11}}e^{2i\zeta_kT_+}\\
&=\frac{[D_K(t;\zeta_k)]_{22}}{[D_K(t;\zeta_k)]_{21}}e^{2i\zeta_kt}
-\frac{[D_K(T_+;\zeta_k)]_{12}}{[D_K(T_+;\zeta_k)]_{11}}e^{2i\zeta_kT_+}\\
&=\frac{\psi_2(t;\zeta_k)}{\phi_2(t;\zeta_k)}-
\frac{\psi_1(T_+;\zeta_k)}{\phi_1(T_+;\zeta_k)}\equiv0.
\end{align*}
Therefore, 
$\omega_j(\zeta_k)+\pi_j(\zeta_k)\exp[{2i\zeta_k(t-T_+)}]=0$, so that
$\lim_{\zeta\rightarrow\zeta_k}
{\left[\omega_j(\zeta_k)+\pi_j(\zeta_k)\exp[{2i\zeta(t-T_+)}]\right]}/{(\zeta-\zeta_k)}
=2i(t-T_+)\pi_j(\zeta_k)\exp[{2i\zeta_k(t-T_+)}]$. Therefore,
\begin{equation}
\begin{split}
&\vv{\mathcal{I}}(\zeta;t)+\vv{\mathcal{J}}(\zeta;t)\\
&=-\sum_{k=1}^K\frac{\vs{\pi}(\zeta_k;t)}{\zeta-\zeta_k}
\left[e^{2i\zeta (t-T_+)}-e^{2i\zeta_k(t-T_+)}\right]\\
&\quad-\sum_{k=1}^K\frac{\vs{\pi}(\zeta_k^*;t)}{\zeta-\zeta_k^*}
\left[e^{2i\zeta (t-T_+)}-e^{2i\zeta_k^* (t-T_+)}\right],
\end{split}
\end{equation}
is a vector valued function analytic for all $\zeta\in\field{C}$.

Finally, the scattering coefficients for the windowed potential 
$q^{(\sqcap)}(t;T_-,T_+)$ work out to be
${a^{(\sqcap)}}(\zeta)=1+\mathcal{I}^*_1(\zeta^*;-T_-)+\mathcal{J}^*_1(\zeta^*;-T_-)$
and 
${b^{(\sqcap)}}(\zeta)=-[\mathcal{I}_2(\zeta;-T_-)+\mathcal{J}_2(\zeta;-T_-)]e^{2i\zeta T_-}$
for $\zeta\in{\field{C}}$. The discrete spectrum can be computed by first
computing the zeros of ${a^{(\sqcap)}}(\zeta)$ (using methods developed for
analytic functions~\cite{DL1967,KB2000}) which gives the eigenvalues and evaluating
${b^{(\sqcap)}}(\zeta)$ at the eigenvalues gives the norming constant.

We conclude this section by demonstrating that the scattering coefficients
obtained above are functions of exponential type: Setting $t=-T_-$
we have, for $|\zeta|>\max_k|\zeta_k|$,
\begin{equation}
\begin{split}
&\left\|\vv{\mathcal{I}}(\zeta;t)+\vv{\mathcal{J}}(\zeta;t)\right\|\\
&\leq\sum_{k=1}^K\frac{\|\vs{\pi}(\zeta_k;-T_-)\|}{2|\zeta_k|}
\left[e^{-4T\Im\zeta}+e^{4T\Im\zeta_k}\right]\\
&\quad+\sum_{k=1}^K\frac{\|\vs{\pi}(\zeta_k^*;-T_-)\|}{2|\zeta_k|}
\left[e^{-4T\Im\zeta}+e^{-4T\Im\zeta_k}\right].
\end{split}
\end{equation}
From here it is straightforward to conclude that 
$a^{(\sqcap)}(\zeta)$ and $b^{(\sqcap)}(\zeta)$ satisfy an estimate of the
form~\eqref{eq:ab-estimate-result}.

\subsection{Conserved quantities and spectral width}
Consider the Fourier spectrum of the multi-soliton potential denoted by
\begin{equation}
Q(\xi)=\int q(t)e^{-i\xi t}dt.
\end{equation}
For convenience, we introduce the notation
\begin{equation}
\langle\xi^n\rangle
=\frac{\frac{1}{2\pi}\int |Q(\xi)|^2\xi^nd\xi}{\frac{1}{2\pi}\int |Q(\xi)|^2d\xi}
\equiv\frac{\frac{1}{2\pi}\int |Q(\xi)|^2\xi^nd\xi}{\|q\|_2^2},
\end{equation}
for moments in the Fourier domain. Let us observe that the following quantities can 
be expressed entirely in terms of the eigenvalues: 
\begin{align}
C_0&=\|q\|_2^2 = 4\sum_{k}\Im{\zeta_k},\\
C_1&=-\int\partial_tq(t)q^*(t)dt=4i\sum_{k}\Im{\zeta^2_k},\\
C_2&=\int\left[|q(t)|^4-\partial_tq^*(t)\partial_tq(t)\right]dt
=-\frac{16}{3}\sum_{k}\Im{\zeta^3_k}.
\end{align}
These quantities do not evolve as the pulse
propagates along the fiber. Further, from the first moment
\begin{equation}
\begin{split}
\langle\xi\rangle & =\frac{\frac{1}{2\pi}\int |Q(\xi)|^2\xi d\xi}{\|q\|^2_2}
=\frac{\int i\partial_tq(t) q^*(t)dt}{\|q\|^2_2}\\
&=-\frac{iC_1}{C_0},
\end{split}
\end{equation}
and the second moment
\begin{equation}
\begin{split}
\langle\xi^2\rangle
&=\frac{\frac{1}{2\pi}\int |Q(\xi)|^2\xi^2 d\xi}{\|q\|^2_2}
=\frac{\int[i\partial_tq(t)] [i\partial_tq(t)]^*dt}{\|q\|^2_2}\\
&=-\frac{C_2}{C_0}+\frac{1}{C_0}\int |q(t)|^4dt,
\end{split}
\end{equation}
we obtained the variance 
$\langle\Delta\xi^2\rangle=\langle\xi^2\rangle -\langle\xi\rangle^2$ as
\begin{equation}
\begin{split}
\langle\Delta\xi^2\rangle
&=\frac{1}{C_0}\int |q(t)|^4dt+\frac{C_1^2}{C_0^2}-\frac{C_2}{C_0}\\
&\leq \|q\|^2_{\infty}+\frac{C_1^2}{C_0^2}-\frac{C_2}{C_0}.
\end{split}
\end{equation}
This quantity characterizes the width of the Fourier spectrum. Note that the biquadratic 
integral must be computed numerically. However, $\|q\|_{\infty}$ can be computed
in a straightforward manner: From~\eqref{DT-iter-pot}, we have 
$\|q_j\|_{\infty} \leq \|q_{j-1}\|_{\infty}+2\Im(\zeta_j)$, we have
\begin{equation}
\|q_K\|_{\infty} \leq 2\sum_{k=1}^K\Im(\zeta_j),
\end{equation}
which yields
\begin{equation}
\langle\Delta\xi^2\rangle\leq \frac{C^2_0}{2}+\frac{C_1^2}{C_0^2}-\frac{C_2}{C_0}.
\end{equation}
Note that this inequality holds irrespective of how the pulse evolves as it
propagates along the fiber. 

Now, turning to the windowed multi-solitons and denoting the
conserved quantities of the windowed signal by $C_j^{(\sqcap)}$ for
$j=0,1,2,\ldots$, we have
\begin{equation}
\langle\Delta\xi^2\rangle^{(\sqcap)}\leq\|q^{(\sqcap)}\|^2_{\infty}
+\left(\frac{C^{(\sqcap)}_1}{C^{(\sqcap)}_0}\right)^2
-\frac{C^{(\sqcap)}_2}{C^{(\sqcap)}_0}.
\end{equation}
Note that as the pulse evolves $\|q^{(\sqcap)}\|_{\infty}$ may not remain
bounded by $\|q\|_{\infty}$ as it does at the initial point, i.e.,
$x=0$. The conserved quantities for the windowed potential can be obtained from 
the asymptotic expansion of $\log[a^{(\sqcap)}(\zeta)]$ as $|\zeta|\rightarrow\infty$ while keeping
$\zeta\in\field{C}_+$. To this end, let
\begin{equation}
{a^{(\sqcap)}}(\zeta)\sim 1
+\frac{a^{(\sqcap)}_1}{2i\zeta}
+\frac{a^{(\sqcap)}_2}{(2i\zeta)^2}
+\frac{a^{(\sqcap)}_3}{(2i\zeta)^3}+\ldots
\end{equation}
as $|\zeta|\rightarrow\infty$ in $\field{C}_+$. The coefficients introduced
above can be explicitly stated as
\begin{multline}
(2i)^{-j}a^{(\sqcap)}_j=
\sum_{k=1}^K\pi_1(\zeta_k^*)^*e^{4i\zeta_kT}\zeta_k^{j-1}\\
+\sum_{k=1}^K\pi_1(\zeta_k)^*e^{4i\zeta^*_kT}(\zeta^*_k)^{j-1}.
\end{multline}
Observing that $C^{(\sqcap)}_j$ are defined as
\begin{equation}
\log {a^{(\sqcap)}}(\zeta)\sim\sum_{j=0}^{\infty}\frac{C^{(\sqcap)}_{j}}{(2i\zeta)^{j+1}},
\end{equation}
the conserved quantities work out to be
\begin{equation}
\begin{split}
C^{(\sqcap)}_0&=a^{(\sqcap)}_1,\\
C^{(\sqcap)}_1&=a^{(\sqcap)}_2-\frac{(a^{(\sqcap)}_1)^2}{2},\\
C^{(\sqcap)}_2&=a^{(\sqcap)}_3-a^{(\sqcap)}_1a^{(\sqcap)}_2 + \frac{(a^{(\sqcap)}_1)^3}{3}.
\end{split}
\end{equation}

\section{Examples}
In the following, we treat a simple example of a doubly-truncated
$1$-soliton in order to demonstrate how to use the recipe provided in this
article. Similar treatment for a $2$-soliton potential is provided 
in~\cite{V2017NS_PREs1}. Further, we present a general example of a
doubly-truncated $6$-soliton where the procedure outlined in this article 
must be implemented numerically. Note that there are no explicit expressions
provided in this article for arbitrary $K$; however, this does not limit our
ability to compute them to machine precision thanks to the iterative Darboux
transformation procedure\footnote{The aforementioned iterative scheme and the recipe
provided in this article requires some care in implementation in order to avoid
arithmetic overflow/underflow of floating point operations. This discussion is
not central to the understanding of the main results presented in this article 
and is, therefore, being
omitted.}.

\subsection{One Soliton}
Consider a $1$-soliton potential with the discrete spectrum $(\zeta_1,b_1)$
where $\zeta=\xi_1+i\eta_1$. The
Darboux matrix can be easily worked out as
\begin{equation}
D_1(t;\zeta)= \zeta \sigma_0-
\begin{pmatrix}
\frac{|\beta_{0}|^2\zeta_1+\zeta_1^*}{1+|\beta_{0}|^2} 
&\frac{(\zeta_1-\zeta_1^*)\beta_{0}}{1+|\beta_{0}|^2}\\
\frac{(\zeta_1-\zeta_1^*)\beta^*_{0}}{1+|\beta_{0}|^2}
&\frac{\zeta_1+\zeta_1^*|\beta_{0}|^2}{1+|\beta_{0}|^2}
\end{pmatrix},
\end{equation}
where $\beta_{0}(t;\zeta_1, b_1) = - (1/b_{1})e^{-2i\zeta_1t}$. Let
$2T=T_++T_-$ and define
$Z_+=1/\beta_0(T_+)$ and $Z_-=\beta_0(-T_-)$ so that
$|Z_{\pm}|=|b_1|^{\mp1}e^{-2\eta_1 T_{\pm}}$. Now
\[
{P}_1(\xi;-T_-)=
-\frac{4\eta_1^2Z_{+}Z_{-}}{\Xi}
\frac{1}{(\xi-\zeta_1)(\xi-\zeta^*_1)}
\]
where $\Xi=(1+|Z_+|^2)(1+|Z_{-}|^2)$ so that
\begin{equation}
\begin{split}
a^{(\sqcap)}(\zeta)&=1+
\frac{2i\eta_1Z^*_{+}Z^*_{-}}{\Xi}\\
&\quad\times\left[\frac{e^{4i\zeta T}-e^{4i\zeta^*_1 T}}{\zeta-\zeta^*_1}
-\frac{e^{4i\zeta T}-e^{4i\zeta_1 T}}{\zeta-\zeta_1}\right].
\end{split}
\end{equation}
The $b$-coefficient can be computed as follows. Observing
\[
{P}_2(\xi;-T_-)=
-\frac{2i\eta_1Z_{+}}{\Xi}
\frac{[(\xi-\zeta_1)+(\xi-\zeta^*_1)|Z_{-}|^2]}{(\xi-\zeta_1)(\xi-\zeta^*_1)},
\]
we have
\begin{multline}
b^{(\sqcap)}(\zeta) = \frac{2i\eta_1b_1|Z_-|^2}{\Xi}
\frac{\left(e^{-2i(\zeta-\zeta_1)T_+}-e^{2i(\zeta-\zeta_1)T_-}\right)}{\zeta-\zeta_1}\\
+\frac{2i\eta_1|Z_+|^2}{b^*_1\Xi}
\frac{\left(e^{-2i(\zeta-\zeta^*_1)T_+}-e^{2i(\zeta-\zeta^*_1)T_-}\right)}{\zeta-\zeta^*_1}.
\end{multline}
Using the asymptotic expansion of $\log[a^{(\sqcap)}(\zeta)]$ as
$|\zeta|\rightarrow\infty$ in $\field{C}_+$, it is also straightforward to workout
\begin{equation}
\|q^{(\sqcap)}\|^2_2=4\eta_1\frac{1-|Z_-|^2|Z_+|^2}{(1+|Z_-|^2)(1+|Z_+|^2)}.
\end{equation}
To the leading order in $|Z_{\pm}|$, the eigenvalue of the windowed $1$-soliton is
given by
\begin{equation}
\zeta_1^{(\sqcap)} \approx \zeta_1-2i\eta_1(|Z_-|^2+|Z_+|^2),
\end{equation}
and the norming constant given by
\begin{equation}
b_1^{(\sqcap)} \approx b_1+4\eta_1b_1(T_-|Z_-|^2-T_+|Z_+|^2).
\end{equation}

\begin{figure}[t!]
\centering
\includegraphics[scale=1.0]{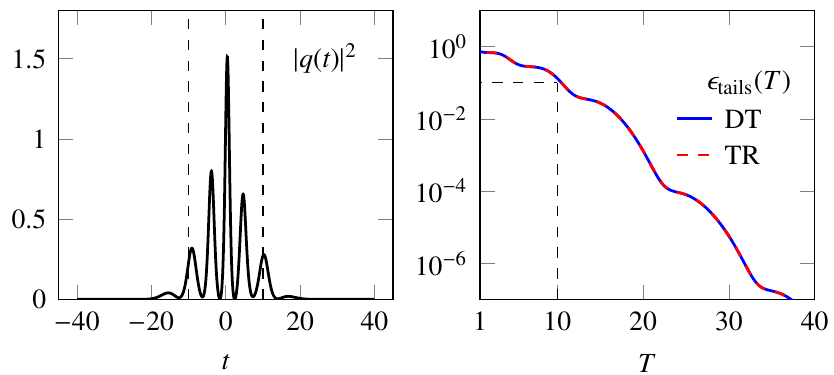}%
\caption{\label{fig:ex2a}The figure shows the $6$-soliton potential (left) corresponding to 
$\mathfrak{S}_6$ (see Fig.~\ref{fig:ex2b}) where the dashed lines mark the truncation points $\pm 10$. The
fraction of the total energy in the tails $\epsilon_{\text{tails}}(T)$ as a
function of the truncation point $\pm T$ is plotted on the right 
(TR: trapezoidal rule).}
\end{figure}
\begin{figure}[t!]
\centering
\includegraphics[scale=1.0]{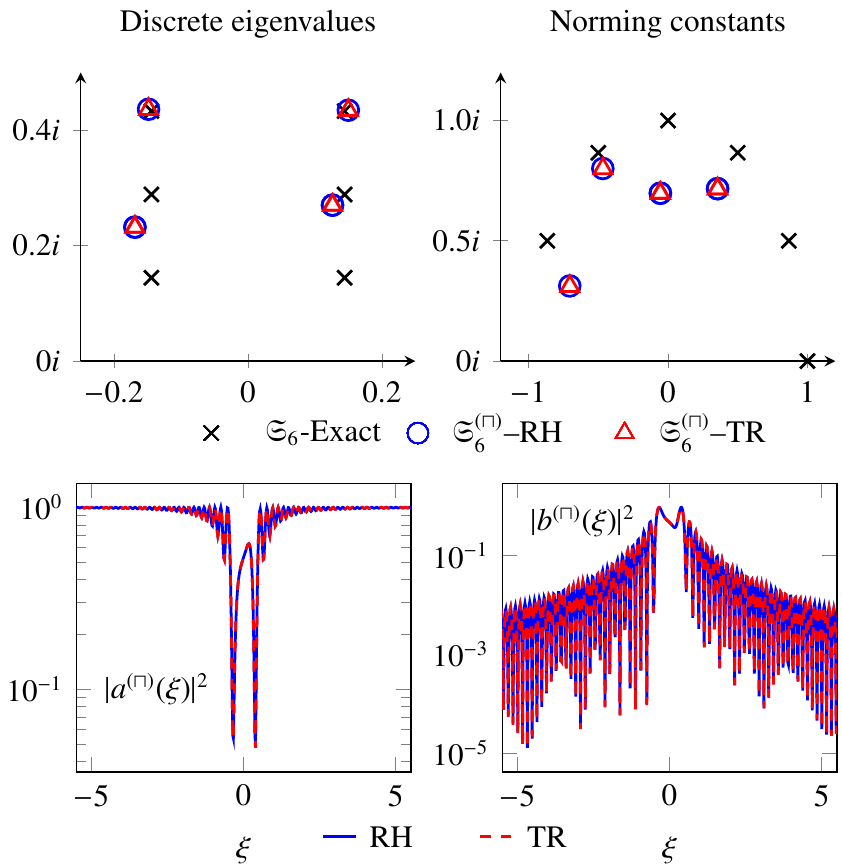}%
\caption{\label{fig:ex2b}The top row shows the discrete spectrum $\mathfrak{S}_6$ of a 
$6$-soliton as well as that of its windowed version, i.e.,
$\mathfrak{S}^{(\sqcap)}_6$. The scattering coefficients of windowed $6$-soliton
is plotted in the bottom row. A comparison is made between our method
(labeled with `RH') and a numerical method of solving the ZS-problem (labeled
with `TR').}
\end{figure}

\subsection{6-Soliton Case}
For the second example,
we choose a complex vector 
$\vs{\lambda}\equiv(\lambda_1,\ldots,\lambda_6)=(\pm 1+1i,\,\pm1+2i,\,\pm1+3i)$ and $b_k
=e^{i(\pi/6)(k-1)},\,k=1,\ldots,6$. The eigenvalues are then taken to be
$\zeta_k=\lambda_k/\kappa$ where $\kappa=2\sqrt{\sum_{k=1}^6\Im\lambda_k}$. The windowed
potential is given by $q^{(\sqcap)}(t;T)=\theta(T^2-t^2)q(t)$ where $T=10$
(see~Fig.~\ref{fig:ex2a}). The energy in the tails is $\epsilon_{\text{tail}}(10)\approx10\%$.
For the sake of comparison, we solve the scattering problem for
$q^{(\sqcap)}(t;T)$ using the (exponential) trapezoidal rule (TR) proposed
in~\cite{V2017INFT1} with $2^{14}$ samples. The method proposed in this letter
is labeled with `RH' in order to signify the fact that an RH-problem is solved
exactly to obtain the scattering coefficients. The discrete spectrum for the windowed potential
`$\mathfrak{S}^{(\sqcap)}_6$--RH' as depicted in Fig.~\ref{fig:ex2b}
is determined numerically from the functional form of $a^{(\sqcap)}(\zeta)$ and
$b^{(\sqcap)}(\zeta)$ obtained above while 
`$\mathfrak{S}^{(\sqcap)}_6$--TR' is computed numerically as
in~\cite{V2017INFT1}. Evidently, results in Fig.~\ref{fig:ex2b} confirm the validity of the method 
proposed in this letter.  
\section{Conclusion}
To conclude, we have discussed a method to solve the Zakharov-Shabat scattering
problem for the doubly-truncated scattering potential in terms of the Jost solutions of the
original potential using the standard techniques of Riemann-Hilbert problems. Exploiting the 
rational structure of the Darboux matrix, it
was possible to obtain the scattering coefficients for the doubly-truncated multi-soliton
potentials. Significance of
this result lies in the fact that on account of the unbounded support of
multi-solitons, windowing is a practical necessity and an important design
issue in optical communication which can now be addressed quite efficiently.
Furthermore, it is interesting to note that the results presented in this article 
may enable us to determine optimal values for various design parameters for
$K$-solitons via the solution of a nonlinear optimization problem. These aspects
will be explored in a future publication.


\providecommand{\noopsort}[1]{}\providecommand{\singleletter}[1]{#1}%

\appendix
\section{Scattering coefficients of compactly supported potentials}
It is well know that Jost solutions and, consequently, the scattering coefficients for 
compactly supported potentials are entire
functions of the spectral parameter $\zeta$, i.e. analytic for all
$\zeta\in\field{C}$~\cite{AKNS1974,AS1981}. Let us consider the scattering
coefficients in the following. It turns out that, for compactly supported potentials, the
scattering coefficients are analytic functions of exponential
type (see~\cite{Boas1954} for properties of such functions) in
$\field{C}$. A simple proof of this statement is provided below. Now, given that a 
doubly-truncated scattering potential is compactly
supported, the method proposed in the article must yield scattering
coefficients that are of exponential type in $\field{C}$.

Introducing the ``local'' scattering coefficients $a(t;\zeta)$ and $b(t;\zeta)$
such that $\vs{\phi}(t;\zeta)=(a(t;\zeta)e^{-i\zeta t}, b(t;\zeta)e^{i\zeta t})^{\tp}$, the scattering
problem in~\eqref{eq:zs-prob} reads as 
\begin{equation}\label{eq:ODE-AB-coeffs}
\begin{split}
&\partial_{t}a(t;\zeta)=q(t)b(t;\zeta)e^{2i\zeta t},\\
&\partial_{t}b(t;\zeta)=r(t)a(t;\zeta)e^{-2i\zeta t}.
\end{split}
\end{equation}
Let $\Omega=[-T_-,T_+]$ where $T_{\pm}\geq0$. The 
initial conditions for the Jost solution $\vs{\phi}$
are: $a(-T_-;\zeta) = 1$ and  $b(-T_-;\zeta) = 0$. The scattering 
coefficients can be directly obtained from these functions as
$a(\zeta)=a(T_+;\zeta)$ and $b(\zeta)=b(T_+;\zeta)$. The following estimate
establishes that $a(\zeta)$ and $b(\zeta)$ are of exponential type in $\field{C}$: 
\begin{theorem}\label{thm:ab-estimate}
Let $q\in\fs{L}^{1}$ with support in $\Omega$ and set $\kappa=\|q\|_{\fs{L}^1(\Omega)}$.
Let $f(\zeta)$ denote either $[a(\zeta)-1]e^{-2i\zeta T_+}$ or $b(\zeta)$; then
the estimate
\begin{equation}\label{eq:ab-estimate-result}
    |f(\zeta)|\leq
\begin{cases}
    Ce^{2T_+\Im{\zeta}},&\zeta\in\ovl{\field{C}}_+,\\
    Ce^{-2T_-\Im{\zeta}},&\zeta\in{\field{C}}_-.
\end{cases}
\end{equation}
holds for $C=\|\vv{D}\|\cosh\kappa$ where $\vv{D}=(\kappa^2/2,\kappa)^{\tp}$.
\end{theorem}
\begin{proof}
Let us define for convenience the modified Jost solution
\begin{equation}
\wtilde{\vv{P}}(t;\zeta) = \vs{\phi}(t;\zeta)e^{i\zeta t}-\begin{pmatrix}1\\0\end{pmatrix}
    =\begin{pmatrix}a(t;\zeta)-1\\b(t;\zeta)e^{2i\zeta t}\end{pmatrix},
\end{equation}
so that
$\wtilde{\vv{P}}(T_+;\zeta)e^{-2i\zeta T_+}
=([a(\zeta)-1]e^{-2i\zeta T_+},b(\zeta))^{\tp}$. The 
system of equations in~\eqref{eq:ODE-AB-coeffs} can be transformed into a set of 
Volterra integral equations of the second kind for $\wtilde{\vv{P}}(t;\zeta)$:
\begin{equation}\label{eq:volterra}
\wtilde{\vv{P}}(t;\zeta)=\vs{\Phi}(t;\zeta)
    +\int_{\Omega}\mathcal{K}(t,y;\zeta)\wtilde{\vv{P}}(y;\zeta)dy,
\end{equation}
where $\vs{\Phi}(t;\zeta)=(\Phi_1,\Phi_2)^{\tp}\in\field{C}^2$ with 
\begin{equation}
\begin{split}
\Phi_1(t;\zeta)&=\int_{-T_-}^{t}q(z)\Phi_2(z;\zeta)dz,\\
\Phi_2(t;\zeta)&=\int_{-T_-}^{t}r(y)e^{2i\zeta(t-y)}dy,
\end{split}
\end{equation}
and the Volterra kernel 
$\mathcal{K}(x,y;\zeta)=\diag(\mathcal{K}_1,\mathcal{K}_2)\in\field{C}^{2\times2}$
is such that
\begin{equation}\label{eq:volterra-kernel}
\begin{split}
&\mathcal{K}_1(x,y;\zeta) = r(y)\int_{y}^{x}q(z)e^{2i\zeta(z-y)}dz,\\
&\mathcal{K}_2(x,y;\zeta) = q(y)\int_{y}^{x}r(z)e^{2i\zeta(x-z)}dz,
\end{split}
\end{equation}
with $\mathcal{K}(x,y;\zeta)=0$ for $y>x$. Now, the proof can be obtained using
the same method as in~\cite{AKNS1974}. For fixed
$\zeta\in\ovl{\field{C}}_+$, let $\OP{K}$ denote the Volterra integral operator
in \eqref{eq:volterra} corresponding to the kernel $\mathcal{K}(x,y;\zeta)$ such
that
\begin{multline}
\OP{K}[\wtilde{\vv{P}}](t;\zeta)=\int_{\Omega}\mathcal{K}(t,y;\zeta)\wtilde{\vv{P}}(y;\zeta)dy\\
=\int_{-T_-}^tdz\int_{-T_-}^zdy
\begin{pmatrix}
q(z)r(y)e^{2i\zeta(z-y)}\wtilde{{P}}_1(y;\zeta)\\
q(y)r(z)e^{2i\zeta(t-z)}\wtilde{{P}}_2(y;\zeta)
\end{pmatrix}.
\end{multline}
Consider the $\fs{L}^{\infty}(\Omega)$-norm~\cite[Chap.~9]{GLS1990} of $\OP{K}$ given by
\begin{equation}
\|\OP{K}\|_{\fs{L}^{\infty}(\Omega)}=\esssup_{t\in\Omega}\int_{\Omega}\|\mathcal{K}(t,y;\zeta)\|dy,
\end{equation}
so that $\|\OP{K}\|_{\fs{L}^{\infty}(\Omega)}\leq\kappa^2/2$~\cite{AKNS1974}. 
The resolvent $\OP{R}$ of this operator exists and is given by the Neumann series
$\OP{R}=\sum_{n=1}^{\infty}\OP{K}_n$ where
$\OP{K}_n=\OP{K}\circ\OP{K}_{n-1}$ with $\OP{K}_{1}=\OP{K}$. It can also be
shown using the methods in~\cite{AKNS1974} that
$\|\OP{K}_n\|_{\fs{L}^{\infty}(\Omega)}\leq{\kappa^{2n}}/{(2n)!}$, 
yielding the estimate 
$\|\OP{R}\|_{\fs{L}^{\infty}(\Omega)}\leq [\cosh(\kappa)-1]$. Therefore, for any
$\vs{\Phi}(t;\zeta)\in\fs{L}^{\infty}(\Omega;\field{C}^2)$, the relationship
$\wtilde{\vv{P}}(t;\zeta)=\vs{\Phi}(t;\zeta)+\OP{R}[\vs{\Phi}](t;\zeta)$
implies, for $\zeta\in\ovl{\field{C}}_+$, 
\begin{equation}\label{eq:estimate-p-uh}
\|\wtilde{\vv{P}}(t;\zeta)\|_{\fs{L}^{\infty}(\Omega)}\leq 
\cosh(\kappa)\|\vs{\Phi}(t;\zeta)\|_{\fs{L}^{\infty}(\Omega)}.
\end{equation}
The result for $\ovl{\field{C}}_+$ in \eqref{eq:ab-estimate-result} follows 
from the observation that, for $\zeta\in\ovl{\field{C}}_+$, 
$\|\vs{\Phi}(t;\zeta)\|_{\fs{L}^{\infty}(\Omega)}\leq\|\vv{D}\|$ 
where $\vv{D}=({\kappa^2}/{2},\kappa)^{\tp}$.
Therefore, $C$ can be chosen to be $\|\vv{D}\|\cosh\kappa$. For 
the case $\field{C}_-$ of \eqref{eq:ab-estimate-result}, we consider
$\wtilde{\vv{P}}_-(t;\zeta)=\wtilde{\vv{P}}(t;\zeta)e^{-2i\zeta t}$ so that
$\wtilde{\vv{P}}_-(T_+;\zeta)
=([a(\zeta)-1]e^{-2i\zeta T_+},b(\zeta))^{\tp}$. The 
Volterra integral equations then reads as
$\wtilde{\vv{P}}_-(t;\zeta)$:
\begin{equation}\label{eq:volterra-lh}
\wtilde{\vv{P}}_-(t;\zeta)=\vs{\Phi}_-(t;\zeta)
    +\int_{\Omega}\mathcal{K}_-(t,y;\zeta)\wtilde{\vv{P}}_-(y;\zeta)dy,
\end{equation}
where $\vs{\Phi}_-(t;\zeta)=\vs{\Phi}(t;\zeta)e^{-2i\zeta t}\in\field{C}^2$ 
and the Volterra kernel 
$\mathcal{K}_-(x,y;\zeta)=\diag(\mathcal{K}^{(-)}_1,\mathcal{K}^{(-)}_2)\in\field{C}^{2\times2}$
is such that
\begin{equation}\label{eq:volterra-kernel-lh}
\begin{split}
&\mathcal{K}^{(-)}_1(x,y;\zeta) = r(y)\int_{y}^{x}q(z)e^{-2i\zeta(x-z)}dz,\\
&\mathcal{K}^{(-)}_2(x,y;\zeta) = q(y)\int_{y}^{x}r(z)e^{-2i\zeta(z-y)}dz,
\end{split}
\end{equation}
with $\mathcal{K}_-(x,y;\zeta)=0$ for $y>x$. Using the approach outlined above,
it is possible to show that, for $\zeta\in\field{C}_-$, 
$\|\wtilde{\vv{P}}_-(t;\zeta)\|_{\fs{L}^{\infty}(\Omega)}\leq 
\cosh(\kappa)\|\vs{\Phi}_-(t;\zeta)\|_{\fs{L}^{\infty}(\Omega)}$.
The result for the case $\zeta\in\field{C}_-$ in \eqref{eq:ab-estimate-result}
then follows from the observation that 
$\|\vs{\Phi}_-(t;\zeta)\|_{\fs{L}^{\infty}(\Omega)}\leq \|\vv{D}\|e^{-2\Im(\zeta)T_-}$ for
$\zeta\in\field{C}_-$.
\end{proof}


\end{document}


\title{Exact Solution of the Zakharov-Shabat Scattering Problem for 
Doubly-Truncated Multi-Soliton Potentials: Supplementary Material}


\author{V.~Vaibhav}
\email{v.k.vaibhav@tudelft.nl}

\affiliation{Delft Center for Systems and Control, 
Delft University of Technology, Mekelweg 2. 2628 CD Delft, 
The Netherlands}
\date{\today}



\maketitle
\section{Some bounds on the energy in the tails}\label{sec:lin-system-DT}
The unknown Darboux matrix coefficients introduced in the article can 
also be obtained as the solution of a
linear system~\cite{NM1984,V2017INFT1}. From 
$\vs{\phi}_K(t;\zeta_k) = b_{k}\vs{\psi}_K(t;\zeta_k)$, we have 
\begin{equation}\label{eq:linear-eq-Darboux}
D_{K}(t;\zeta_k,\mathfrak{S}_K)
[\vs{\phi}_0(t;\zeta_k)-b_{k}\vs{\psi}_0(t;\zeta_k)]=0.
\end{equation}
Note that $\vs{\phi}_0(t;\zeta_k) - b_{k}(t)\vs{\psi}_0(t;\zeta_k)\neq0$ on
account of the fact that $\zeta_k$ is not an
eigenvalue of the seed potential. Define the Vandermonde matrix
$\mathcal{V}=\{V_{jk}\}_{K\times K}$ where 
\[
V_{jk}=\zeta_j^{k},\quad j=1,2,\ldots,K,\,k=0,1,\ldots,K-1,
\] 
and the diagonal matrix $\Gamma=\text{diag}(\gamma_1,\gamma_2,\ldots,\gamma_K)$.
Let the vectors
\begin{equation}
\vv{f}=\begin{pmatrix}
\zeta_1^K\\
\zeta_2^K\\
\vdots\\
\zeta_{K}^K\\
\end{pmatrix},\,\,\vv{g}=\Gamma\vv{f}=\begin{pmatrix}
\zeta_1^K\gamma_1\\
\zeta_2^K\gamma_2\\
\vdots\\
\zeta_{K}^K\gamma_K\\
\end{pmatrix},
\end{equation}
where 
\begin{equation}\label{eq:gama-DT}
\gamma_k = \frac{\phi_2^{(0)}(t;\zeta_k) - b_{k}\psi_2^{(0)}(t;\zeta_k)}
{\phi_1^{(0)}(t;\zeta_k)-b_{k}\psi_1^{(0)}(t;\zeta_k)}
=-b_ke^{2i\zeta_k t}.
\end{equation}
The unknown Darboux coefficients can be put into the vector form
\begin{equation}
\vv{D}_0=\begin{pmatrix}
d_0^{(0,K)}\\
d_0^{(1,K)}\\
\vdots\\
d_0^{(K-1,K)}
\end{pmatrix},\,\,
\vv{D}_1=\begin{pmatrix}
    d_1^{(0,K)}\\
    d_1^{(1,K)}\\
\vdots\\
    d_1^{(K-1,K)}
\end{pmatrix}.
\end{equation}
The $2K$ linear system of equations~\eqref{eq:linear-eq-Darboux} can be written as
\begin{equation}\label{eq:DT-linear-sys}
-\begin{pmatrix}
\vv{f}\\
\vv{g}^*
\end{pmatrix}=\begin{pmatrix}
\mathcal{V} & \Gamma \mathcal{V}\\
\Gamma^* \mathcal{V}^* & -\mathcal{V}^*
\end{pmatrix}
\begin{pmatrix}
\vv{D}_0\\
\vv{D}_1
\end{pmatrix}.
\end{equation}
Let $\eta_{\text{min}}=\min_k(\Im{\zeta_k})$ and 
\begin{equation}
\varpi_+=\max_k(|b_k|),\quad\varpi_-=\max_k\left(|b_k|^{-1}\right).
\end{equation}
Consider the case $t=T_+>0$.
Denoting the spectral norm of matrices by $\|\cdot\|_s$, we have 
\begin{equation}\label{eq:Gamma-norm-Tp}
\|\Gamma\|_s=\max_{k}\left(|b_k|e^{-2\Im{\zeta_k}T_+}\right)
\leq\varpi_+e^{-2\eta_{\text{min}}T_+},
\end{equation}
so that $\|\Gamma\|_s\rightarrow0$ as $T_+\rightarrow\infty$. Define
$\vv{D}_0(\infty)=-\mathcal{V}^{-1}\vv{f}$. Putting 
$\vv{D}_0=\vv{D}_0(\infty)+\wtilde{\vv{D}}_0$, we have
\begin{align*}
&\wtilde{\vv{D}}_0=-M \vv{D}_1,\\
&\vv{D}_1 =
\left(1+M^*M\right)^{-1}M^*\left[\vv{D}_0(\infty)-\vv{D}^*_0(\infty)\right],
\end{align*}
where $M=\mathcal{V}^{-1}\Gamma\mathcal{V}$. Let 
$\kappa(\mathcal{V})=\|\mathcal{V}^{-1}\|_s\|\mathcal{V}\|_s$ denote the condition
number of $\mathcal{V}$, then $\|M\|_s\leq\kappa\|\Gamma\|_s$. If $T_+$ is such that 
$\kappa\|\Gamma\|_s<1$, then
\begin{align*}
&\|\wtilde{\vv{D}}_0\|_2\leq \kappa\|\Gamma\|_s\|\vv{D}_1\|_2,\\
&\|\vv{D}_1\|_2 \leq
\frac{2\kappa\|\Gamma\|_s}{\left(1-\kappa^2\|\Gamma\|^2_s\right)}\|\Im[\vv{D}_0(\infty)]\|_2.
\end{align*}
Note that ${d}^{(K-1, K)}_{0}(\infty)=-\sum_{k=1}^K\zeta_k$, therefore,
\begin{equation}\label{eq:q-tail-estimate-p}
\mathcal{E}^{(+)}_K(T_+)\leq\frac{2\kappa^2\|\Gamma\|^2_s}
{\left(1-\kappa^2\|\Gamma\|^2_s\right)}\|\Im[\vv{D}_0(\infty)]\|_2.
\end{equation}

Next we consider the case $t=-T_-<0$. The linear system~\eqref{eq:DT-linear-sys}
can be written as
\begin{equation}
-\begin{pmatrix}
\Gamma^{-1}\vv{f}\\
\vv{f}^*
\end{pmatrix}=\begin{pmatrix}
\Gamma^{-1}\mathcal{V} & \mathcal{V}\\
\mathcal{V}^* & -(\Gamma^{-1})^*\mathcal{V}^*
\end{pmatrix}
\begin{pmatrix}
\vv{D}_0\\
\vv{D}_1
\end{pmatrix}.
\end{equation}
Observing
\[
\|\Gamma^{-1}\|_s=\max_{k}\left(\frac{e^{-2\Im{\zeta_k}T_-}}{|b_k|}\right)
\leq\varpi_-e^{-2\eta_{\text{min}}T_-},
\]
implies $\|\Gamma^{-1}\|_s\rightarrow0$ as $T_-\rightarrow\infty$. Define
$\vv{D}_0(-\infty)=-[\mathcal{V}^{-1}\vv{f}]^*$. As before, putting 
$\vv{D}_0=\vv{D}_0(-\infty)+\wtilde{\vv{D}}_0$, we have
\begin{align*}
&\wtilde{\vv{D}}_0=(M^{-1})^* \vv{D}_1,\\
&\vv{D}_1
=-\left(1+M^{-1}(M^{-1})^*\right)^{-1}M\left[\vv{D}_0(-\infty)-\vv{D}^*_0(-\infty)\right].
\end{align*}
If $T_-$ is such that $\kappa\|\Gamma^{-1}\|_s<1$, then
\begin{align*}
&\|\wtilde{\vv{D}}_0\|_2\leq \kappa\|\Gamma^{-1}\|_s\|\vv{D}_1\|_2,\\
&\|\vv{D}_1\|_2 \leq
\frac{2\kappa\|\Gamma^{-1}\|_s}{\left(1-\kappa^2\|\Gamma^{-1}\|^2_s\right)}\|\Im[\vv{D}_0(-\infty)]\|_2.
\end{align*}
Note that ${d}^{(K-1, K)}_{0}(-\infty)=-\sum_{k=1}^K\zeta^*_k$, therefore,
\begin{equation}\label{eq:q-tail-estimate-m}
\mathcal{E}_K^{(-)}(-T_-)\leq\frac{2\kappa^2\|\Gamma^{-1}\|^2_s}
{\left(1-\kappa^2\|\Gamma^{-1}\|^2_s\right)}\|\Im[\vv{D}_0(-\infty)]\|_2.
\end{equation}

\begin{rem} 
Define 
\begin{equation}
\begin{split}
T_+^{(0)}&=\log(\kappa\varpi_+)/2\eta_{\text{min}},\\
T_-^{(0)}&=\log(\kappa\varpi_-)/2\eta_{\text{min}},
\end{split}
\end{equation}
then, $T_+>T_+^{(0)}$ and $T_->T_-^{(0)}$ ensures that the
estimates~\eqref{eq:q-tail-estimate-p} and~\eqref{eq:q-tail-estimate-m} hold, respectively.
Now, the effect of propagation can also be incorporated by plugging in the
$x$-dependence of the norming constant in~\eqref{eq:Gamma-norm-Tp}. For fixed
$x\in[0, L]$, we seek $T_+\in\field{R}$ such that $\|\Gamma\|_s<\kappa^{-1}$, where
\[
\|\Gamma\|_s
=\max_{k}\left(|b_k|e^{-2\eta_k(T_+-4\xi_k x)}\right).
\]
Let $k_{\text{min}}$ be such that
$\Im(\zeta_{k_{\text{min}}})=\eta_{\text{min}}$, then putting 
$\xi_{\text{min}}=\Re(\zeta_{k_{\text{min}}})$, we have
\[
\|\Gamma\|_s
\leq\varpi_+e^{-2\eta_{\text{min}}(T_+-4\xi_{\text{min}}x)}.
\]
Therefore, choosing $T_+-4\xi_{\text{min}}x>T_+^{(0)}$ 
ensures that the estimate~\eqref{eq:q-tail-estimate-p} holds.
Using similar arguments as above, it follows that choosing $T_-+4\xi_{\text{min}}x>T_-^{(0)}$ 
ensures the validity of estimate~\eqref{eq:q-tail-estimate-m}.
\end{rem}

Now, in the design of $K$-soliton pulses, if the tolerance for the fraction of
total energy in the tails is $\epsilon_{\text{tails}}$, then the domain
$[-T_-,T_+]$ must be chosen such that
\[
\frac{\mathcal{E}^{(+)}_K(T_+)+\mathcal{E}^{(-)}_K(-T_-)}{4\sum_{k=1}^K\Im\zeta_k}
\leq\epsilon_{\text{tails}}.
\]
If we choose to satisfy the equality above, then one has to solve a nonlinear
equation for $T_+,T_-$. For the sake of simplicity, we let $T=\max(T_-, T_+)$. The
inequalities obtained above can be used to compute an upper bound for $T$ as
follows: set $X=\exp(-4\eta_{\text{min}}T_{\text{max}})$, then 
\begin{equation}
\frac{2\kappa^2\varpi_+^2X}{1-\kappa^2\varpi_+^2X}+
\frac{2\kappa^2\varpi_-^2X}{1-\kappa^2\varpi_-^2X}
=\frac{4\epsilon_{\text{tails}}\sum_k\Im\zeta_k}{\|\Im[\mathcal{V}^{-1}\vv{f}]\|_2}.
\end{equation}
A good estimate for $T_{\text{max}}$ is 
\begin{equation}
T_{\text{max}}
=\frac{1}{4\eta_{\text{min}}}\log\left[
\frac{\|\Im[\mathcal{V}^{-1}\vv{f}]\|_2\kappa^2(\varpi_+^2+\varpi_-^2)}
{2\epsilon_{\text{tails}}\sum_k\Im\zeta_k}\right],
\end{equation}
provided the RHS is positive. Finally, a search algorithm such as the \emph{bisection method}
can be used to obtain the true value of $T$ by choosing a bracketing interval
of the form $[0,T_{\text{max}}]$.

\section{Example: Symmetric multi-solitons}
Reflection symmetry can be considered by putting $\lambda=-\zeta^*$ and 
introducing $\vv{w}(s;\lambda) = \vv{v}^*(-s;-\lambda^*)$ so that the ZS-problem
reads as
\begin{equation*}
\vv{w}_s = -i\lambda\sigma_3\vv{w}+{U}(-s)\vv{w}.
\end{equation*}
Denote the Jost solutions of the
new system by $\vs{\Psi}(s;\lambda)$, $\ovl{\vs{\Psi}}(s;\lambda)$
(first kind) and $\vs{\Phi}(s;\lambda)$, $\ovl{\vs{\Phi}}(s;\lambda)$ (second
kind); then
\begin{align*}
&\vs{\Psi}(s;\lambda)=\sigma_1{\vs{\phi}}^*(-s;-\lambda^*),\,
\ovl{\vs{\Psi}}(s;\lambda)=-\sigma_1\ovl{\vs{\phi}}^*(-s;-\lambda^*),\\
&\vs{\Phi}(s;\lambda)=\sigma_1{\vs{\psi}}^*(-s;-\lambda^*),\,
\ovl{\vs{\Phi}}(s;\lambda)=-\sigma_1\ovl{\vs{\psi}}^*(-s;-\lambda^*).
\end{align*}
Let $B_k$ denote the norming constant for $U(-s)$, then it follows that
$B_k=1/b_k^*$. Therefore, for symmetric profiles, $U(s) = U(-s)$, we have $B_k\equiv
b_k=1/b_k^*$ or $|b_k|=1$. Let $A(\lambda)$ and $B(\lambda)$ be the scattering coefficients for 
the new system, then
\begin{align*}
A(\lambda)=\OP{W}(\vs{\Phi},{\vs{\Psi}})=a^*(-\lambda^*),\\
B(\lambda)=\OP{W}(\ovl{\vs{\Psi}},\vs{\Phi}) = \ovl{b}^*(-\lambda^*),
\end{align*}
Therefore, if $\zeta_k$ is an eigenvalue then so is $-\zeta^*_k$. If $U(s)$ is
assumed to be symmetric even as it evolves according to the NSE, then
$\zeta_k=i\eta_k$ where $\eta_k>0$.
The constraint on the Darboux matrix (assuming $K$ is even) can be stated as
\begin{equation}
D(s;\lambda,\mathfrak{S}_K)=\sigma_1D^*(-s;-\lambda^*,\mathfrak{S}_K)\sigma_1,
\end{equation}
which translates to 
\begin{equation}
\begin{split}
&d_0^{(k,K)}(s)=(-1)^kd_0^{(k,K)}(-s),\\
&d_1^{(k,K)}(s)=(-1)^{k+1}d_1^{(k,K)}(-s). 
\end{split}
\end{equation}
Given that the potential is symmetric, we choose the truncation points also
symmetrically so that $T=T_-=T_+$. These simplifying assumption allow
us to study the $2$-soliton case analytically.
\subsubsection{Symmetric $2$-soliton}
Continuing from the
Sec.~\ref{sec:lin-system-DT}, we observe that
\begin{equation} 
\begin{split} 
&d_0^{(0,2)}(\infty) = -\eta_1\eta_2,\\
&d_0^{(1,2)}(\infty)=-i(\eta_1+\eta_2).
\end{split} 
\end{equation} 
Let the discrete spectrum be 
\begin{equation}
\mathfrak{S}_2=\{(i\eta_1,e^{i\theta_1}),\,(i\eta_2,e^{i\theta_2})|
\,\eta_j,\theta_j\in\field{R},\,j=1,2\}.
\end{equation} 
Introducing $X_j=e^{-2\eta_j T}$ for $j=1,2$, the solution to the 
linear system in~\eqref{eq:DT-linear-sys} can be stated as
\begin{widetext}
\begin{align}
&\tilde{d}_0^{(0,2)} = 2\eta_1\eta_2(\eta_1+\eta_2)
\left[(\eta_1+\eta_2)(X_1^2+X_2^2) 
-2X_1X_2\left(\eta_1e^{-i(\theta_1-\theta_2)}+\eta_2e^{i(\theta_1-\theta_2)}\right)\right]/\Xi,\\
%
&\tilde{d}_0^{(1,2)}=2i(\eta_1+\eta_2)\left[
(\eta_1 X_1 -\eta_2 X_2)^2 
    + (\eta_1-\eta_2)^2 X_1^2 X_2^2    
+\eta_1\eta_2\{(X_1+X_2)^2- 4X_1 X_2\cos(\theta_1-\theta_2)\}\right]/\Xi\\
&d_1^{(0,2)} = -2\eta_1\eta_2(\eta_1^2-\eta_2^2)
\left[e^{-i\theta_1}X_1(1-X_2^2)-e^{-i\theta_2}X_2(1-X_1^2)\right]/\Xi,\\
&d_1^{(1,2)} = 2i(\eta_1^2-\eta_2^2)
\left[\eta_1e^{-i\theta_1}X_1(1+X_2^2)-\eta_2e^{-i\theta_2}X_2(1+X_1^2)\right]/\Xi,
\end{align}
where
\begin{equation}\label{eq:Xi-sym-soliton}
\begin{split}
\Xi&= \left[(\eta_1-\eta_2)^2(1+X^2_1X^2_2)+(\eta_1+\eta_2)^2 (X^2_1+X^2_2)
    -8X_1 X_2\eta_1\eta_2\cos(\theta_1-\theta_2)\right]\\
&= \left[(\eta_1-\eta_2)^2(1-X_1X_2)^2+(\eta_1+\eta_2)^2 (X_1-X_2)^2     
    +4X_1 X_2(\eta_1^2+\eta_2^2-2\eta_1\eta_2\cos(\theta_1-\theta_2))\right].
\end{split}
\end{equation}
The potential works out to be
\begin{equation}
q(t) = 2id_1^{(1,2)} = \frac{-4(\eta_1^2 - \eta^2_2)}{\Xi}\left[\eta_1e^{-i\theta_1} X_1(1+X_2^2) 
-\eta_2e^{-i\theta_2} X_2(1+X_1^2)\right].
\end{equation}
Given that the potential is symmetric, it suffices to consider the energy in the
tail $[T,\infty)$. From the relation $\mathcal{E}^{(+)}(T)=2i\tilde{d}_0^{(1,2)*}$, we have
\begin{equation}\label{eq:Eplus}
\begin{split}
\mathcal{E}^{(+)}(T)&=\frac{4 (\eta_1 + \eta_2)}{\Xi}
[(\eta_1 X_1 -\eta_2 X_2)^2 + (\eta_1-\eta_2)^2 X_1^2 X_2^2
+\eta_1\eta_2((X_1+X_2)^2- 4X_1 X_2\cos(\theta_1-\theta_2))]\\
&= 2 (\eta_1 + \eta_2)-\frac{4 (\eta_1 + \eta_2)}{\Xi}
[(\eta^2_1-\eta^2_2) (X_2^2-X_1^2)+(\eta_1-\eta_2)^2 (1-X_1^2 X_2^2)].
\end{split}
\end{equation}
Without the loss generality, we may assume that $\eta_1>\eta_2$ so that
$X^2_2>X^2_1$ and $1>X_1^2X_2^2$. Therefore, from~\eqref{eq:Xi-sym-soliton}
and~\eqref{eq:Eplus}, it follows that maximum value of the energy in the tails is achieved when
$\theta_1=\theta_2+2n\pi$ where $n\in\field{Z}$. Next, using the the symmetry
relations for the Darboux matrix coefficients, we may write
\begin{equation}
P_1(\xi;-T)=-\frac{\left[d_1^{(1,2)*}(T)\xi+d_1^{(0,2)*}(T)\right]
\left[d_1^{(1,2)}(T)\xi-d_1^{(0,2)}(T)\right]}
{(\xi+i\eta_1)(\xi-i\eta_1)(\xi+i\eta_2)(\xi+i\eta_2)},
\end{equation}
and
\begin{equation}
P_2(\xi;-T)=-\frac{\left[d_1^{(1,2)*}(T)\xi+d_1^{(0,2)*}(T)\right]
\left[\xi^2-d_0^{(1,2)*}(T)\xi+d_0^{(0,2)*}(T)\right]}
{(\xi+i\eta_1)(\xi-i\eta_1)(\xi+i\eta_2)(\xi+i\eta_2)}.
\end{equation}
This yields the scattering coefficients as
\begin{equation}
\begin{split}
a^{(\sqcap)}(\zeta)&=1-\frac{\pi^*_1(i\eta_1)}{(\zeta+i\eta_1)}
\left[e^{4i\zeta T}-X_1^{-2}\right]
-\frac{\pi^*_1(-i\eta_1)}{(\zeta-i\eta_1)}\left[e^{4i\zeta T}-X_1^2\right]\\
&\qquad-\frac{\pi^*_1(i\eta_2)}{(\zeta+i\eta_2)}\left[e^{4i\zeta T}-X_2^{-2}\right]
-\frac{\pi^*_1(-i\eta_2)}{(\zeta-i\eta_2)}\left[e^{4i\zeta T}-X_2^2\right],
\end{split}
\end{equation}
and
\begin{equation}
\begin{split}
b^{(\sqcap)}(\zeta)&=
\frac{\pi_2(i\eta_1)}{(\zeta-i\eta_1)}\left[e^{-2i\zeta T}-X_1^{-2}e^{2i\zeta T}\right]
+\frac{\pi_2(-i\eta_1)}{(\zeta+i\eta_1)}\left[e^{-2i\zeta T}-X_1^2e^{2i\zeta T}\right]\\
&\quad+\frac{\pi_2(i\eta_2)}{(\zeta-i\eta_2)}\left[e^{-2i\zeta T}-X_2^{-2}e^{2i\zeta T}\right]
+\frac{\pi_2(-i\eta_2)}{(\zeta+i\eta_2)}\left[e^{-2i\zeta T}-X_2^2e^{2i\zeta T}\right].
\end{split}
\end{equation}
Putting $Y=(\eta_1+\eta_2)/(\eta_1-\eta_2)$, the change $\Delta\eta_j$ in the
eigenvalues, to the leading
order in $X_j$, works out to be
\begin{equation}
\begin{split}
\frac{\Delta\eta_1}{\eta_1+\eta_2}&=-2Y(Y+1)X_1^2+2(Y^2-1)X_1X_2\cos(\theta_1-\theta_2),\\
\frac{\Delta\eta_2}{\eta_1+\eta_2}&=-2Y(Y-1)X_2^2+2(Y^2-1)X_1X_2\cos(\theta_1-\theta_2),
\end{split}
\end{equation}
and the corresponding change $\Delta b_j$ in the norming constants work out to be
\begin{equation}\label{eq:trunc-eigs}
\begin{split}
\frac{\Delta b_1}{\eta_1+\eta_2} &=
4i(Y^2-1)T\left(1-\frac{1}{4\eta_1T}\right)e^{i\theta_1}X_1X_2\sin(\theta_1-\theta_2),\\
\frac{\Delta b_2}{\eta_1+\eta_2} &=
4i(Y^2-1)T\left(1-\frac{1}{4\eta_2T}\right)e^{i\theta_2}X_1X_2\sin(\theta_2-\theta_1).
\end{split}
\end{equation}
The limiting forms in~\eqref{eq:trunc-eigs} indicate that the eigenvalue with smaller imaginary part suffers 
greater change as a result of truncation to $[-T,T]$. 
\end{widetext}

\bibliography{NS}